\documentstyle[12pt]{article}

\newcommand{\EQ}{\begin{equation}}
\newcommand{\EN}{\end{equation}}

\newcommand{\bear}{\begin{eqnarray}}
\newcommand{\ear}{\end{eqnarray}}

\begin{document}

\topmargin 0pt
\oddsidemargin 5mm
\newcommand{\NP}[1]{Nucl.\ Phys.\ {\bf #1}}
\newcommand{\PL}[1]{Phys.\ Lett.\ {\bf #1}}
\newcommand{\NC}[1]{Nuovo Cimento {\bf #1}}
\newcommand{\CMP}[1]{Comm.\ Math.\ Phys.\ {\bf #1}}
\newcommand{\PR}[1]{Phys.\ Rev.\ {\bf #1}}
\newcommand{\PRL}[1]{Phys.\ Rev.\ Lett.\ {\bf #1}}
\newcommand{\MPL}[1]{Mod.\ Phys.\ Lett.\ {\bf #1}}
\newcommand{\JETP}[1]{Sov.\ Phys.\ JETP {\bf #1}}
\newcommand{\TMP}[1]{Teor.\ Mat.\ Fiz.\ {\bf #1}}

\renewcommand{\thefootnote}{\fnsymbol{footnote}}

\newpage
\setcounter{page}{0}
\begin{titlepage}
\begin{flushright}
UFSCARF-TH-94-23
\end{flushright}
\vspace{0.5cm}
\begin{center}
{\large Bethe ansatz solution of the $Osp(1|2n)$ invariant spin chain}\\
\vspace{1cm}
\vspace{1cm}
{\large M.J.  Martins  } \\
\vspace{1cm}
{\em Universidade Federal de S\~ao Carlos\\
Departamento de F\'isica \\
C.P. 676, 13560~~S\~ao Carlos, Brasil}\\
\end{center}
\vspace{1.2cm}

\begin{abstract}
We have applied the  analytical Bethe ansatz approach in order to solve the
$Osp(1|2n)$
invariant magnet. By using the Bethe ansatz equations we have calculated the
ground
state energy and the low-lying dispersion relation. The finite size properties
indicate
that the model has a central charge $c=n$.
\end{abstract}
\vspace{.2cm}
\vspace{.2cm}
\centerline{February 1995}
\end{titlepage}

\renewcommand{\thefootnote}{\arabic{footnote}}
\setcounter{footnote}{0}

\newpage
In the last past years many
hierarchies of exactly vertex models have been found \cite{REV}. Of
particular interest are those invariant by superalgebras $Sl(n|m)$ and
$Osp(n|2m)$ in which
the associated Boltzmann weights satisfy a graded version of the Yang-Baxter
relation \cite{KS,BA}. An interesting example of a system in statistical
mechanics which
can be realized in terms of the $Sl(n|m)$ series is the Perk-Shultz model
\cite{PK,DV}.
Although the Bethe ansatz properties of the $Sl(n|m)$ model have
been examined in different contexts \cite{KS,DV,KO} in the literature, not much
is
known concerning the exact solution of the $Osp(n|2m)$ hierarchy. In fact, only
recently that the isotropic $Osp(1|2)$ chain \cite{MA} (see also ref.\cite{KS})
and
$U_q Osp(2|2)$ models \cite {MAS} have been solved in detail by the Bethe
ansatz approach. In this sense, we believe that it is important to make an
effort
in order to find the exact solution of a more general class of $Osp(n|2m)$
invariant
magnets.

The purpose of this Letter is to present the analytical Bethe ansatz solution
of the
$Osp(1|2n)$ spin chain for $any$ value of $n$. This extends  previous results
found
by the author in the simplest case of the $Osp(1|2)$ magnet.
The ground state energy, the low-lying dispersion relation and the
associated central charge are also computed by using the corresponding
Bethe ansatz equations.

The $Osp(1|2n)$ chain is defined by the following Hamiltonian,
\EQ
H= -J\sum_{i=1}^{L} [ P^g_{i,i+1} +\frac{2}{2n+1} E^g_{i,i+1} ]
\EN
where periodic boundary conditions are assumed, and $L$ is the number of sites
of the
lattice. Here we are interested in the antiferromagnetic regime $J>0$ of
Hamiltonian (1). The symbols $P^g_{i,i+1}$ and $E^g_{i,i+1}$ stand for the
graded
permutation operator \cite{KS} and the $Osp(1|2n)$ Temperely-Lieb invariant
\cite{MA,CH},
respectively. More precisely \cite{MA,CH}, we have the matrix elements
$(E^g_{i,i+1})_{ab}^{cd}= \alpha_{ab} \alpha_{cd}^{st}$ where the matrix
$\alpha$ is
defined by
\EQ
\alpha=\left( \begin{array}{cc}
	1 &   O_{1X2n} \\
	O_{2nX1} & \left( \begin{array}{cc} O_{nXn} & I_{nXn} \\
	-I_{nXn} & O_{nXn} \\ \end{array} \right)\\
	\end{array}
	\right)
\EN
where $I_{aXa}$ ($O_{aXa}$) is the identity (null) $aXa$ matrix.

The problem of diagonalization of Hamiltonian (1) is equivalent to that of
finding
the Bethe ansatz solution of the corresponding vertex system. For instance,
the model (1) is obtained as the logarithmic derivative of the following
Transfer matrix $T(\lambda, \eta)$ at $\lambda=0$,
\EQ
T(\lambda,\eta) =Tr_0[ {\cal L}_{0L}(\lambda,\eta) \cdots {\cal
L}_{01}(\lambda,\eta)]
\EN
where $ {\cal L}(\lambda,\eta)_{ab}^{cd} = (-1)^{p(a)p(b)}
R(\lambda,\eta)_{ab}^{cd}$
\cite{KS,MA}, and the indices $p(a)$ and $0$ denote the Grasman parities
\footnote{ In the
case of the $BFF \cdots F$ ( one boson and 2n fermions)
grading we have $p(1)=0$, $p(i)=1$ $i=2,
\cdots ,2n+1$ .} and $(2n+1)X(2n+1)$ auxiliary space, respectively. The
operator
$R(\lambda,\eta)$ is the $Osp(1|2n)$ solution of the graded Yang-Baxter
equation given
by
\EQ
(R(\lambda, \eta)_{i,i+1})_{ab}^{cd} = \lambda \delta_{ac} \delta_{bd} +
\eta (P^g_{i,i+1})_{ab}^{cd}
+\frac{\eta \lambda}{(n+1/2)\eta -\lambda}(E^g_{i,i+1})_{ab}^{cd}
\EN

The first step toward the diagonalization of (3) is to perform a redefinition
of the
grading to $F \cdots B \cdots F$ by exchanging the first (bosonic) and the
$(n+1)$-esimo (fermionic) degrees of freedom. As we shall see below, such
canonical
transformation has the advantage of adapting  the vertex
operator $\cal{L} (\lambda,\eta)$ in a more symmetric way,
before acting it on the reference state. The next step is to notice that
the usual ferromagnetic vacuum defined by
\EQ
|0> = \prod_{i}^{L} |0>_i; ~~~ |0>_i= \left( \begin{array}{c} 1 \\ 0 \\
\vdots \\ 0 \end{array} \right )
\EN
is an eigenvector of the Transfer-matrix (3). In fact, the vertex
$\cal{L}(\lambda,\eta)$
has a triangular form when acting on such reference state, namely
\EQ
{\cal L}(\lambda,\eta) |0> = \left ( \begin{array}{ccccc}
\eta -\lambda & * & * & \cdots & * \\
0  & \lambda & * & \cdots & *\\
0  & 0 & \lambda &  \cdots & *\\
\vdots  & \vdots & \vdots &  \vdots & \vdots \\
0 & 0 &  0 & 0 &\frac{\lambda((n-1/2)\eta -\lambda)}{\lambda - (n+1/2)\eta}
\end{array} \right )
\EN
which leads to the following eigenvalue $\Lambda(\lambda)$ of $T(\lambda,\eta)$
\EQ
\Lambda(\lambda)= (\eta-\lambda)^L
+\sum_{j=1}^{2n-1} \lambda^L
+\left [ \frac{\lambda([n-1/2]\eta-\lambda)}{\lambda-(n+1/2)/\eta} \right ]^L
\EN

In accordance to the hypothesis of the analytical Bethe ansatz approach
\cite{RE}, one
now seeks for a more general ansatz of form
\bear
\Lambda(\lambda,\{ \lambda_ j^k\}) =
(\eta-\lambda)^L \prod_{j=1}^{M_1} A(\lambda-\lambda_j^1) +\lambda^L
\sum_{k=1}^{2n-1} \prod_{j=1}^{M_k} B(\lambda,\lambda_j^k,\lambda_j^{k+1})
\nonumber \\
+\left [ \frac{\lambda([n-1/2]\eta-\lambda)}{\lambda-(n+1/2) \eta}
\right ]^L \prod_{j=1}^{M_1} C(\lambda-\lambda_j^1)
\ear
where $A(x)$, $B_j(x)$ and $C(x)$ are some rational functions which can be
fixed by
using the crossing symmetry, unitarity condition and the asymptotic behaviour
of
$\Lambda(\lambda)$. In particular, we have found the following relations
between
these functions
\EQ
C((n+1/2)\eta-x) =A(x),~~ B_{2n-k}((n+1/2)\eta -x)=B_k(x),
{}~ B_1(x)= A(x-\eta/2)A^{-1}(x-\eta)
\EN

Taking into account that the amplitude $A(x)$ satisfies the ``unitarity ''
condition,
$A(x) A(-x)=1$, and from our previous experience with the $Osp(1|2)$ case
\cite{MA},
we can set this function as
\EQ
A(x)= -\frac{x+\eta/2}{x-\eta/2}
\EN
and by using interactively this solution on Eq.(9) we then end up with the
following
result
\bear
\Lambda(\lambda,\{ \lambda_ j^k\}) =
(i-\lambda)^L \prod_{j=1}^{M_1} -\frac{\lambda-\lambda_j^1+i/2}
{\lambda-\lambda_j^1-i/2}
+\lambda^L \left \{ \sum_{k=1}^{n-1}\prod_{j=1}^{M_k}
-\frac{\lambda-\lambda_j^k-i(k+2)/2}{\lambda-\lambda_j^k-ik/2}
\right . \nonumber \\ \left .
\prod_{j=1}^{M_{k+1}}
-\frac{\lambda-\lambda_j^{k+1}-i(k-1)/2}{\lambda-\lambda_j^{k+1}-i(k+1)/2}
 + \prod_{j=1}^{M_n}
\left ( \frac{\lambda-\lambda_j^{n}-i(n+2)/2}
{\lambda-\lambda_j^{n}-i(n)/2} \right )
\left ( \frac{\lambda-\lambda_j^{n}-i(n-1)/2}{\lambda-\lambda_j^{n}-i(n+1)/2}
\right )
\right . \nonumber \\ \left .+
\sum_{k=n+1}^{2n-1}\prod_{j=1}^{M_{2n-k}}
-\frac{\lambda-\lambda_j^{2n-k}-i(k-1)/2}{\lambda-\lambda_j^{2n-k}-i(k+1)/2}
\prod_{j=1}^{M_{2n-k+1}}
-\frac{\lambda-\lambda_j^{2n-k+1}-i(k+2)/2}{\lambda-\lambda_j^{2n-k+1}-ik/2}
\right \}
\nonumber \\ +
\left [ \frac{\lambda([n-1/2]i-\lambda)}{\lambda-[n+1/2]i}
\right ]^L \prod_{j=1}^{M_1}
-\frac{\lambda-\lambda_j^1-i(n+1)}{\lambda-\lambda_j^1-in}
\ear
where we conveniently choose $\eta=i$ due to the scale invariance
$\lambda \rightarrow \eta \lambda$.

Finally, the condition that the residues of $\Lambda(\lambda,\{ \lambda_j^l \})
$ at
$\lambda =\lambda_j^l +l/2 $, $l=1,\cdots,n $ vanishes, then fixes the
following Bethe ansatz
equations for  the variables $\{ \lambda_j^l \}$ \footnote{ Analogously, the
crossing
symmetry ( last two terms of Eq.(11) ) guarantee that the extra poles at
$\lambda =
\lambda_j^{2n+1-l} +l/2$, $l=n+1,\cdots, 2n$ produces the same restriction (12)
for
the set $\{ \lambda_j^l \} $. }

\bear
{\left ( \frac{\lambda_j^1 -i/2}{\lambda_j^1 +i/2}\right )}^{L}
=
(-1)^{L-M_2} \prod_{k=1}^{M_1}
\frac{\lambda_j^1 -\lambda_k^1 -i}{\lambda_j^1-\lambda_k^1 +i}
\prod_{k=1}^{M_2} \frac{\lambda_j^1 -\lambda_k^2 +i/2}{\lambda_j^1-\lambda_k^2
-i/2}
\nonumber \\
\prod_{k=1}^{M_l} \frac{\lambda_j^l -\lambda_k^l -i}{\lambda_j^l-\lambda_k^l
+i}  =
(-1)^{M_{l-1}-M_{l+1}}
\prod_{k=1}^{M_{l+1}} \frac{\lambda_j^{l} -\lambda_k^{l+1} -i/2}
{\lambda_j^{l}-\lambda_k^{l+1} +i/2}
\prod_{k=1}^{M_{l-1}} \frac{\lambda_j^{l} -\lambda_k^{l-1} -i/2}
{\lambda_j^{l}-\lambda_l^{l-1} +i/2},~l=2,\cdots,n-1
\nonumber \\
\prod_{k=1}^{M_n} \left (\frac{\lambda_j^n -\lambda_k^n
-i}{\lambda_j^n-\lambda_k^n +i}
\right ) \left (\frac{\lambda_j^n -\lambda_k^n +i/2}
{\lambda_j^n-\lambda_k^n -i/2} \right )  =
(-1)^{M_{n-1}-M_{n}}
\prod_{k=1}^{M_{n-1}} \frac{\lambda_j^{n} -\lambda_k^{n-1} -i/2}
{\lambda_j^{n}-\lambda_l^{n-1} +i/2}
\ear
where the numbers $M_l$ are related to the many sectors indices $r_l$ of the
theory
by $M_l=L-r_l$. The eigenenergies $E(L)$ of Hamiltonian (1) are parametrized in
terms of the
Bethe ansatz roots $\{\lambda^1_j \}$ by
\EQ
E(L) =-\sum_{j=1}^{M_1} \frac{1}{(\lambda_j^1)^2 +1/4} + L
\EN

An interesting characteristic of these equations is the appearance of the phase
factors
$\pm 1$, distinguishing the behaviour of roots $\{ \lambda_j^l \} $ in the
sectors $r_l$
of  the system. Here we  mention that similar factors have also been found in
the
solution of the Perk-Shultz model \cite {DV}. In particular, for the simplest
case of
the $Osp(1|2)$ chain, such phase plays the role of an index which physical
meaning
is to split the even and odd degrees of freedom present in the theory
\cite{MA}. In any case,
such phase factors will change the logarithm branches of the Bethe ansatz
equations (12) and
therefore they are extremely important in the correct characterization of the
ground
state and the low-lying excitations.

Let us turn to the computation of some properties in the thermodynamic  limit
of
Hamiltonian (1). We have found that the ground state in a given sector $ r_l$
is parametrized
by a set of real roots $\{ \lambda_j^l \} $ of the Bethe ansatz equations. In
this case, by
taking the logarithm of Eqs.(12) we find that
\bear
L \phi_{1/2}(\lambda_j^1) =2 \pi Q_j^1 + \sum_{k=1,k \neq j}^{M_1}
 \phi_{1}(\lambda_j^1-\lambda_k^1) -
\sum_{k=1}^{M_2}\phi_{1/2}(\lambda_j^1-\lambda_k^2)
\nonumber \\
\sum_{k=1,k \neq j}^{M_l}\phi_{1}(\lambda_j^l-\lambda_k^l) +2 \pi Q_j^l =
\sum_{k=1}^{M_{l-1}}\phi_{1/2}(\lambda_j^l-\lambda_k^{l-1})
+ \sum_{k=1}^{M_{l+1}}\phi_{1/2}(\lambda_j^l-\lambda_k^{l+1}),~l=2,\cdots,n-1
\nonumber \\
\sum_{k=1,k \neq j}^{M_n}[\phi_{1}(\lambda_j^n-\lambda_k^n)
-\phi_{1/2}(\lambda_j^n-\lambda_k^n)]
+2 \pi Q_j^n =
\sum_{k=1}^{M_{n-1}}\phi_{1/2}(\lambda_j^n-\lambda_k^{n-1})
\ear
where $\phi_a(x)= 2\arctan(x/a)$ and $Q_j^l$ are the following numbers
characterizing the different
branches of the logarithm
\EQ
Q_j^{l}= -\frac{[L-r_l-1]}{2} +j-1,~~ j=1,2, \cdots, L-r_l
\EN

In the thermodynamic limit, $L \rightarrow \infty $, the roots $\{ \lambda_j^l
\} $ cluster
into a continuous distribution of densities $\rho_l(\lambda)$ satisfying a set
of $n$-coupled
integral equations. This system of integral equations can be solved by
elementary Fourier
techniques, and here we only summarize our results
\EQ
\rho_l(\lambda) =\frac{4}{2n+1} \frac{ \cos[\frac{(2n+1-2l)\pi}{2(2n+1)}]
\cosh[\frac{2\lambda \pi}{(2n+1)}]}{\cosh[\frac{4 \lambda \pi}{2n+1}]
+\cos[\frac{( 2n+1-2l)\pi}{2n+1}]}
\EN

The ground state per particle is calculated by using density $\rho_1(\lambda)$
in Eq.(13), after
replacing the sum by an integral. The final result is
\EQ
e_{\infty} = -\int_{-\infty}^{\infty} \frac{\rho_1(\lambda)}{\lambda^2 +1/4}
d \lambda +1 = 1-\frac{2}{2n+1} \left \{2\ln(2)+
\psi[1/2+1/(2n+1)]-\psi[1/(2n+1)] \right \}
\EN
where $\psi(x)$ is the Euler psi-function. The low-lying excitation over the
ground state are
obtained by the insertion of ``holes '' on the distribution of the numbers
$Q_j^l$. In general,
this generates $n$ branches of excitations and we find that all of them are
gapless. More precisely,
the low-momentum $p$ dispersion relation has the behaviour
\EQ
\epsilon_l(p) = \frac{2 \pi}{2n+1} p , l=1,\cdots,n
\EN
and therefore we find a unique sound velocity $v_s= \frac{2 \pi}{2n+1} $.

To conclude we would like to present some numerical results of the finite size
behaviour
of the ground state of Hamiltonian (1). Such finite size effects can be
explicitly related
to the central charge governing the underlying conformal field theory of this
system. For instance,
by extrapolating the following sequence \cite{CA},
\EQ
\frac{E(L)}{L}= e_{\infty} -\frac{\pi v_s c}{6 L^2}
\EN
we are able to compute the central charge $c$. In order to do that, we have
solved numerically
the Bethe ansatz equations up $L=44$ and by substituting the solution $\{
\lambda_j^1 \}$ in
Eq.(13) we have determined the ground state energy for finite $L$. In Table 1,
we present our
results for the estimatives (19) in the case of $n=2$ and $n=3$. Taking into
account these
results and those from our previous study of the $Osp(1|2)$ chain ($c=1$), we
conjecture that
the underlying central charge is $c=n$. In spite of our numerical results, this
conjecture is
also confirmed by using the analytical method of refs. \cite{DW,RES}, since we
have $n$-nested
Bethe ansatz equations of $real$ roots. One possible physical interpretation of
this result
is as follows. Recalling that this system is made by one bosonic and $2n$
fermionic degrees
of freedom, we may conclude that only the fermions contribute to the central
charge, namely
$c=(2n)/2=n$.

In summary we have presented the Bethe ansatz solution, the ground state and
the low-lying
dispersion relation of the $Osp(1|2n)$ spin chain. Interesting enough, we have
noticed
that the ground state structure resembles much that appearing in the
$O(2n+3)$($B_{n+1}$)
\cite{MA1}  invariant
magnets. We believe that this fact indicates that ( as we have shown for the
$Osp(1|2)$ model \cite{MA}) the $Osp(1|2n)$ chain can be obtained as a peculiar
and $new$
branch limit of the anisotropic $A_{2n-1}^2$ vertex model. Our results together
with those of
ref. \cite{VL} for the conformal anomaly $c$, strongly suggest that the
anisotropic $A_{2n-1}^2$
model has in fact many gapless regimes. We hope to address such questions in
future
publications.

\section*{Acknowledgements}
This work is
supported by CNPq and FAPESP (Brazilian agencies).

\vspace*{0.5cm}
\newpage
\centerline{ \bf Table Captions}
\vspace*{0.5cm}
Table 1. The estimative of the conformal anomaly from equation (19).

\begin{center}
{\bf Table 1}\\
\vspace{0.5cm}
\begin{tabular}{|l|l|l|} \hline
L & $ n=2$ & n=3 \\ \hline \hline
8 & 1.97 108 & 2.85 851\\ \hline
16 & 1.99 703 & 2.97 544 \\ \hline
24 & 2.00 032 & 2.99 170  \\ \hline
32 &  2.00 118 & 2.99 675\\ \hline
40 & 2.00 144  & 2.99 887 \\ \hline
48 & 2.00 152  & 2.99 991 \\ \hline
Extrapolated & 2.00 0(3) & 3.00 0(1)\\ \hline
\end{tabular}
\end{center}

\end{document}